\documentclass[twoside]{dis09}
\usepackage[latin1]{inputenc}
\usepackage[dvips]{graphicx,epsfig,color}
\usepackage{wrapfig,rotating}
\usepackage{amssymb,amsmath,array}

\begin{document}
\title{Ultra-peripheral Au+Au collisions at PHENIX}

\author{M\'at\'e Csan\'ad$^1$ for the PHENIX Collaboration
\vspace{.3cm}\\
1 - E\"otv\"os University - Department of Atomic Physics \\
P\'azm\'any P\'eter s. 1/a, Budapest, H-1117 Hungary}

\maketitle

\begin{abstract}
Ultra-peripheral collisions (UPC) of heavy-ions involve long range electromagnetic interactions at impact parameters twice larger than the nuclear radius, where no nucleon-nucleon collisions occur.
The first measurement of photoproduction of $J/\psi$ and of two-photon production of high-mass $e^+ e^-$ pairs in ultra-peripheral nucleus-nucleus interactions will be presented, using Au+Au data at $\sqrt{s_{_{NN}}}$ = 200 GeV. The measured cross sections at midrapidity are consistent with various theoretical predictions.~\cite{Csanad:DISTalk}
\end{abstract}

\section{Introduction}

The study of photoproduction at hadron colliders has 
attracted an increased interest in recent years~\cite{Baur:2001jj,Bertulani:2005ru,Baltz:2007kq}. 
Electromagnetic interactions can be studied without background from hadronic 
processes in ultra-peripheral collisions without nuclear overlap.  
This study focuses on the measurement of exclusively produced high-mass $e^+ e^-$--pairs
in Au+Au collisions at $\sqrt{s_{_{NN}}}$ = 200 GeV, $Au+Au \rightarrow Au+Au + e^+ e^-$
at midrapidity. The results~\cite{Afanasiev:2009hy} have been obtained with the PHENIX detector~\cite{Adcox:2003zm} at the BNL 
Relativistic Heavy Ion Collider (RHIC). 

The electromagnetic field of a relativistic particle can be represented by 
a spectrum of equivalent photons. This is the Weizsacker-Williams approximation.  
The virtualities of the equivalent photons when the field couples 
coherently to the entire nucleus are restricted by the nuclear form factor 
to $Q^2=\left(\omega^2/\gamma^2+q_\perp^2\right)\lesssim \hbar /R_A^2$, 
where $\gamma$ is the Lorentz factor of the beam and $R_A$ is the nuclear radius. 
At RHIC energies, $\gamma$ = 108 and the maximum photon energy in the 
center-of-mass (lab) system is $\omega_{max}\approx$ 3 GeV corresponding  
to a maximum photon-nucleon center-of-mass energy of  
$W^{max}_{\gamma N} \approx$~34 GeV. 

\begin{figure}[htbp]
\begin{center}
\includegraphics[width=\linewidth]{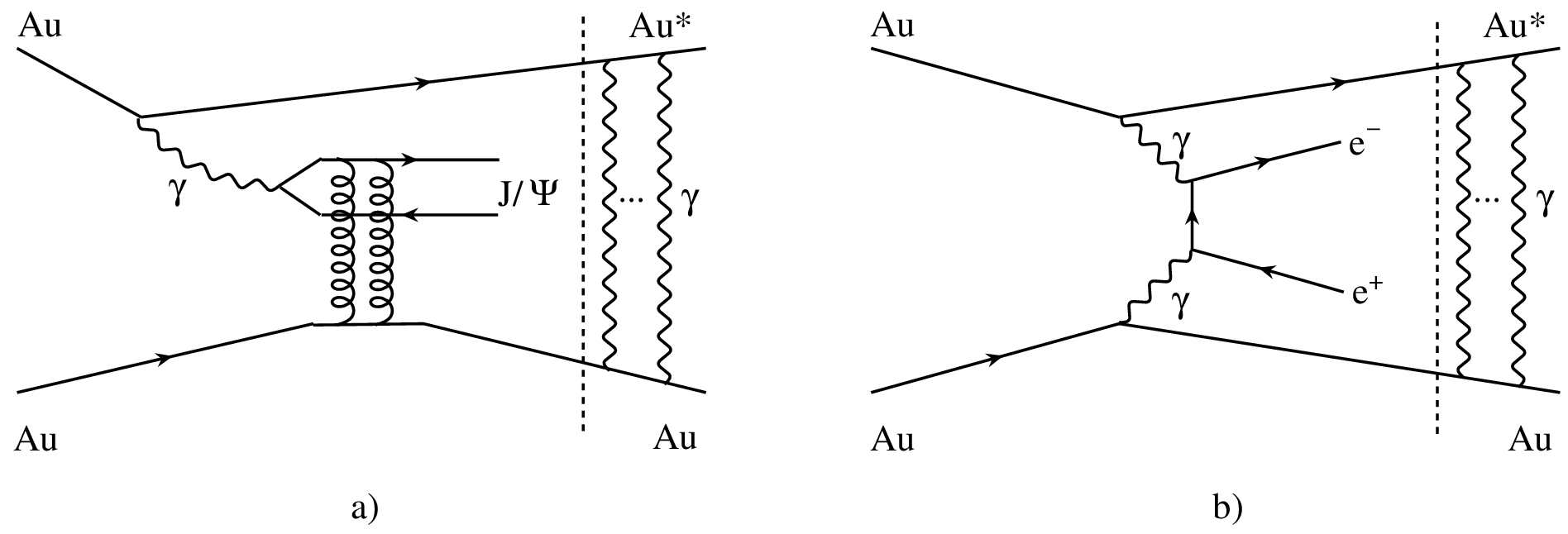}
\end{center}
\caption{Lowest order Feynman diagrams for exclusive photoproduction of $J/\psi$ (left) and 
dielectrons (right) in ultra-peripheral Au+Au collisions. 
The photons to the right of the dashed line are soft photons that may 
excite the nuclei but do not lead to particle production.}
\label{fig:diag_gg_gA}
\end{figure}

The exclusive production of an $e^+ e^-$--pair can proceed either through a 
purely electromagnetic process or through coherent photonuclear production of a vector meson, which decays 
into a dilepton pair. The Feynman diagrams for the two 
leading order processes are shown in Fig.~\ref{fig:diag_gg_gA}. 

The production cross section of vector mesons is a good probe of the nuclear gluon 
distribution, $G_A(x,Q^2)$, as well as of vector-meson dynamics in nuclear 
matter~\cite{Ryskin:1995hz,Frankfurt:1995jw,Frankfurt:2001db}. For $J/\psi$-production,
PHENIX acceptance at $y=0$ corresponds to a mean photon energy
of $\left<E_{\gamma}\right>= 300$~GeV and nuclear Bjorken-$x$ values of 
$x = m_{J/\psi}^2/W_{\gamma A}^2 \approx 1.5 \cdot 10^{-2}$. 

Measurements of coherent photonuclear 
production of the $\rho$ meson~\cite{Adler:2002sc}, as well as $\gamma\,\gamma$ production 
of {\it low-mass} $e^\pm$ pairs~\cite{Adams:2004rz} have been performed by the 
STAR collaboration. The PHENIX analysis presented in Refs.~\cite{d'Enterria:2006ep,Afanasiev:2009hy} and summarized in this paper is the first on 
heavy final states in ultra-peripheral nucleus-nucleus collisions. The cross section for 
$J/\psi$ and $e^+ e^-$ photoproduction are compared with model 
calculations~\cite{Frankfurt:2001db,Klein:1999qj,Strikman:2005ze,Goncalves:2005sn,Ivanov:2007ms}.

\section{Experimental analysis}

The data presented here were collected with the PHENIX detector at RHIC 
during the 2004 Au+Au run at $\sqrt{s_{_{NN}}}$ = 200 GeV (Run-4). The PHENIX detector
is equipped with multi-layer drift chambers (DC) followed by multi-wire proportional chambers 
(PC) with pixel-pad readout. The tracking arms also contain 
Ring-Imaging-\v{C}erenkov (RICH) detectors and 
electromagnetic calorimeters (EMCal) for electron and 
positron identification.

The events used in this analysis were collected by a level-1 Ultra-Peripheral Collision (UPC)
trigger set up for the first time in PHENIX in Run-4 as follows. A veto on coincident signals in both Beam-Beam Counters  (BBC) 
selected exclusive-type events characterised by a large rapidity gap on either side of the central arm. The EMCal and the RICH decetor were used to form a trigger (ERT) to select
events with at least one of the two high-energy $e^\pm$ coming from the $e^+e^-$ pair. Finally at least 30 GeV energy deposited in one or both of the ZDCs was required to 
select events with forward neutron emission.

The total number of events collected by the UPC trigger was 8.5 M, of which 6.7 M satisfied 
standard quality assurance criteria. The useable event sample corresponds to an integrated 
luminosity $ {\mathcal L}_{int} = 141 \pm 12 \;\mu\mbox{b}^{-1}$. 

The following cuts were applied to enhance the sample of genuine $\gamma$-induced events:
\begin{enumerate}
\item A standard offline vertex cut $|vtx_{z}| <$ 30 cm was required
\item Only events with exactly two charged particles were analyzed. This cut
  allows to suppress most of non photoproduction contamination in the UPC trigger.
\item A RICH cut selects $e^\pm$ which fire 2 or more tubes 
separated by the nominal ring radius.
\item Good Track--EMCal matching is also required.
\item An EMCal energy cut ($E_1 >$ 1 GeV $||$ $E_2 >$ 1 GeV) is applied to 
select candidate $e^\pm$ above the ERT trigger threshold.
\item Events with back-to-back $e^+e^-$ candidates (detected in opposite 
arms) were selected.
\end{enumerate}

\section{Results}

\begin{figure}
\begin{center}
\includegraphics[width=0.56\linewidth]{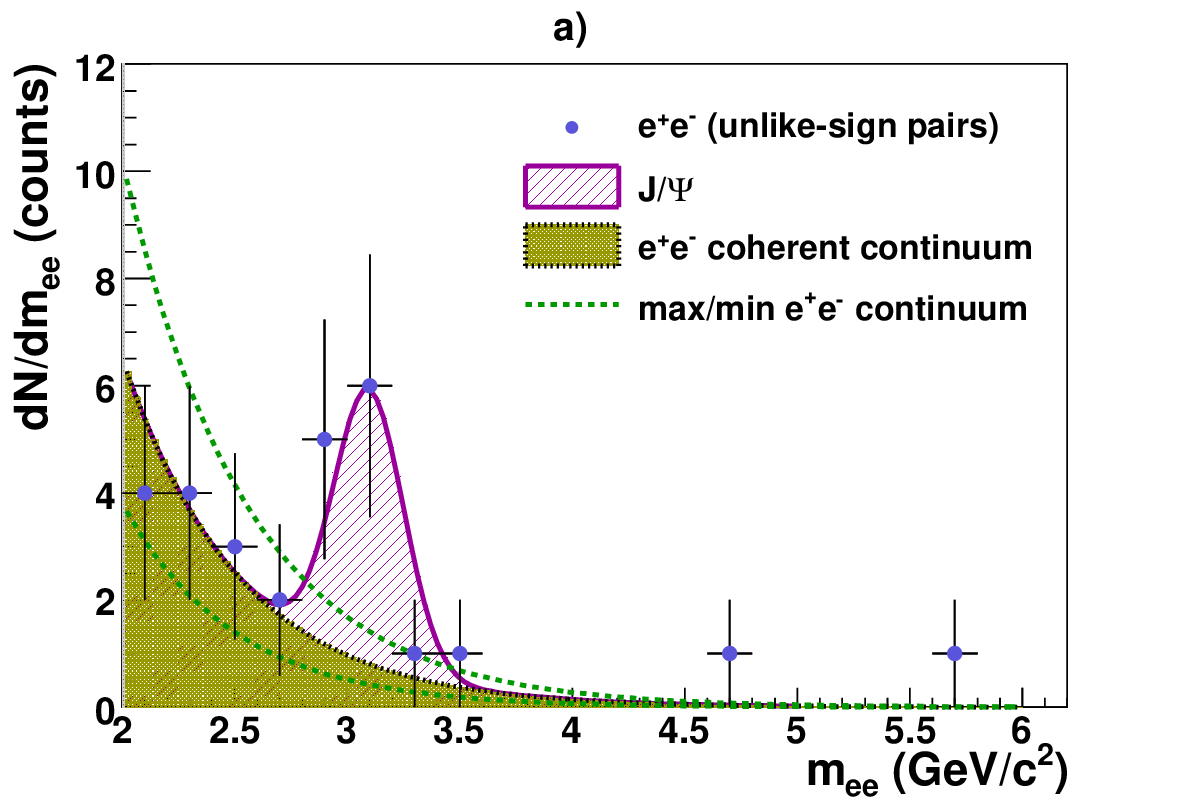} 
\includegraphics[width=0.43\linewidth]{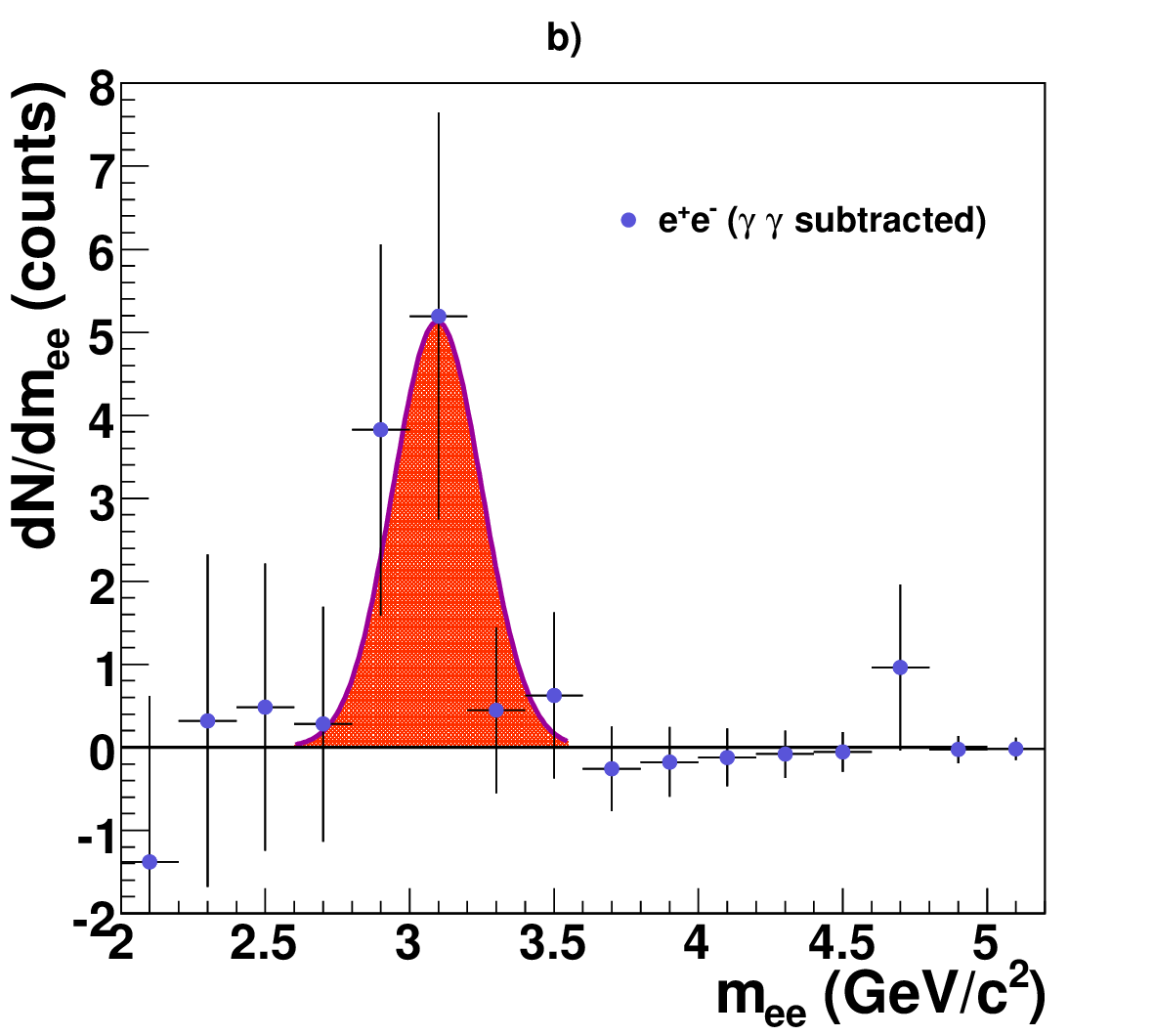} 
\end{center}
\caption{Left: 
(a) Invariant mass distribution of $e^+e^-$ pairs fitted to the
combination of (shaded) a dielectron continuum [exponential distribution]
and (cross hatched) a $J/\psi$ [Gaussian] signal. The two additional dashed
curves indicate the maximum and minimum continuum contributions considered
in this analysis (see text).  (b) $J/\psi$ invariant mass distribution
after subtracting the fitted dielectron continuum signal in (a).
}
\label{fig:minv_ee_jpsi}
\end{figure}

After the above cuts we find 28 events with $e^+e^-$ pairs and none with like-sign pairs for 
$m_{e^{\pm}e^{\pm}}>2$~GeV/c$^2$. The measured $e^+ e^-$ invariant mass distribution for the sample is shown in 
Fig.~\ref{fig:minv_ee_jpsi}~a). This distribution is fitted with a continuum (exponential) 
curve combined with a Gaussian function at the $J/\psi$ peak, as shown by the 
solid curve in Fig.~\ref{fig:minv_ee_jpsi}~a). Simulations based on events 
generated by the {\sc starlight} Monte Carlo~\cite{Klein:1999qj,Baltz:2002pp,Nystrand:2004vn} and processed
through {\sc geant3}~\cite{Geant:321} have shown that 
the shape of the measured continuum contribution is well described by an 
exponential function $dN/dm_{e^+ e^-} = A \cdot e^{ c \, m_{e^+ e^-}}$. Those 
simulations allow us to fix the exponential slope parameter to $c=-1.9 \pm 
0.1$~GeV$^{-1}$c$^2$.  The combined data fit is done with three free 
parameters: the exponential normalisation ($A$), the $J/\psi$ yield and the 
$J/\psi$ peak width (the Gaussian peak position has been fixed at the known 
$J/\psi$ mass of $m_{J/\psi}$~=~3.097~GeV/c$^2$~\cite{Amsler:2008zzb}). The $J/\psi$ and continuum yields and the corresponding statistical errors are 
calculated from the fit and summarized in Table~\ref{tab:results}. Fig.~\ref{fig:minv_ee_jpsi}~b) shows the resulting invariant mass distribution 
obtained by subtracting the fitted exponential curve of the dielectron 
continuum from the total experimental $e^+e^-$ pairs distribution.

Physical cross-sections were obtained after correcting the raw number of signal
counts for the geometrical acceptance of our detector system, and 
 the efficiency losses introduced by the previously described analysis cuts.
Acceptance and efficiency corrections were obtained with a 
full Monte Carlo of the experimental apparatus with realistic 
{\sc starlight} Monte Carlo.
Such a model reproduces well the existing $d^3N/dyd\phi dp_T$ distribution of coherent 
$\rho$ production in UPC Au+Au events measured at RHIC by STAR~\cite{Adler:2002sc}.
The coherent events were simulated in the PHENIX detector using {\sc geant3} 
and passed through the same reconstruction program as the real data.

For $J/\psi$ at midrapidity the differential cross section is calculated as:
\begin{equation}
\left .\frac{d\sigma_{J/\psi + Xn}}{dy}\right|_{|y|<0.35} =  
\frac{1}{\ensuremath{{\it BR}}}\cdot\frac{N_{J/\psi}}{\ensuremath{{\it
      Acc}}\cdot \varepsilon \cdot
\varepsilon_{\ensuremath{\it trigg}}\cdot {\mathcal L}_{int}}\cdot\frac{1}{\Delta y}.
\nonumber
\end{equation}
\noindent
Here $Acc$ is the detector acceptance, $\varepsilon$ is the track reconstruction efficiency, $\varepsilon_{trigg}$
is the trigger efficiency. The integrated luminosity $\mathcal{L}_{int}$ is given in Section 2, $\Delta y$ is the rapidity interval 
of the measurement. These correction factors and corresponding uncertainties are quoted in Ref.~\cite{Afanasiev:2009hy},and BR = 
5.94\% is the known $J/\psi$ dielectron branching ratio~\cite{Amsler:2008zzb}.

For dielectrons at midrapidity ($y$ is the rapidity of the pair) the double differential cross 
section is:
\begin{equation}
 \left .\frac{d^2 \sigma_{e^+ e^- + Xn}}{d y \, d
     m_{e^+e^-}}\right|_{|y|<0.35, \, \Delta m_{e^+e^-}} = 
  \frac{N_{e^+ e^-}}{ \ensuremath{{\it Acc}} \cdot \varepsilon \cdot
    \varepsilon_{trigg} \cdot \mathcal{L}_{int} } \cdot
  \frac{1}{\Delta y} \cdot \frac{1}{\Delta m_{e^+e^-}},
  \nonumber
\end{equation} 
where the factors are defined as for the previous equation.

\begin{table}
  \begin{center}
    \begin{tabular}{|c|c|c|c|} \hline
      $m_{e^+e^-}$ [GeV/c$^2$]  & Yield & Cross-section [$\mu$b/(GeV/c$^2$)]   \\  \hline
      $J/\psi$ peak            & $9.9 \pm 4.1~{\rm(st)} \pm 1.0~{\rm(sy)}$ &
                                 $76  \pm  31~{\rm(st)} \pm  15~{\rm(sy)} $ [$\mu$b]  \\  \hline
      $e^+e^-$ cont. [2.0,2.8] & $13.7\pm 3.7~{\rm(st)} \pm 1.0~{\rm(sy)}$ &
                                 $86  \pm  23~{\rm(st)} \pm  16~{\rm(sy)} $ [{\sc starl.}: $90$] \\ \hline 
      $e^+e^-$ cont. [2.0,2.3] & $7.4 \pm 2.7~{\rm(st)} \pm 1.0~{\rm(sy)}$ &
                                 $129 \pm  47~{\rm(st)} \pm  28~{\rm(sy)}$ [{\sc starl.}: $138$] \\   \hline 
      $e^+e^-$ cont. [2.3,2.8] & $6.2 \pm 2.5~{\rm(st)} \pm 1.0~{\rm(sy)}$ & 
                                 $60  \pm  24~{\rm(st)} \pm  14~{\rm(sy)} $ [{\sc starl.}: $61$] \\ \hline 
    \end{tabular}
  \end{center}
  \caption{    \label{tab:results}
Measured $J/\psi$ and $e^+ e^-$ continuum photoproduction yields and cross-sections at 
midrapidity in ultra-peripheral Au+Au collisions (accompanied with forward 
neutron emission) at $\sqrt{s_{_{NN}}}$~=~200~GeV (obtained from the fit of
the data to an exponential plus Gaussian function) per invariant mass range
are shown in the second column. For the continuum
cross-sections {\sc starlight} predictions are taken from Ref.~\cite{Nystrand:2004vn}.  }
\end{table}

The measured dielectron cross sections at midrapidity are in very good 
agreement with the {\sc starlight} predictions for coherent dielectron 
photoproduction (rightmost column of 
Table~\ref{tab:results})~\cite{Nystrand:2004vn}.

\begin{figure}
\begin{center}
\includegraphics[width=0.46\linewidth]{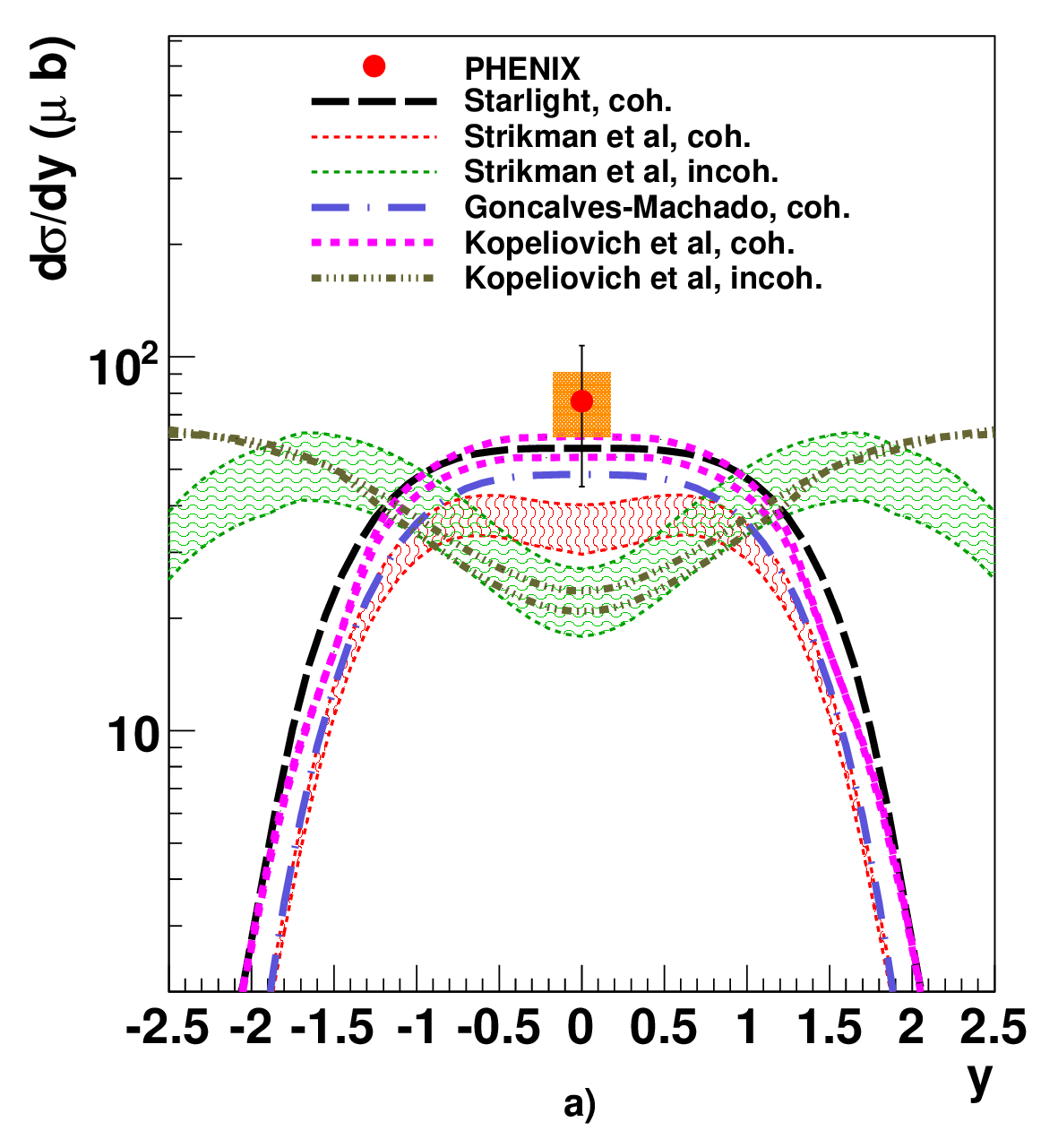}
\includegraphics[width=0.46\linewidth]{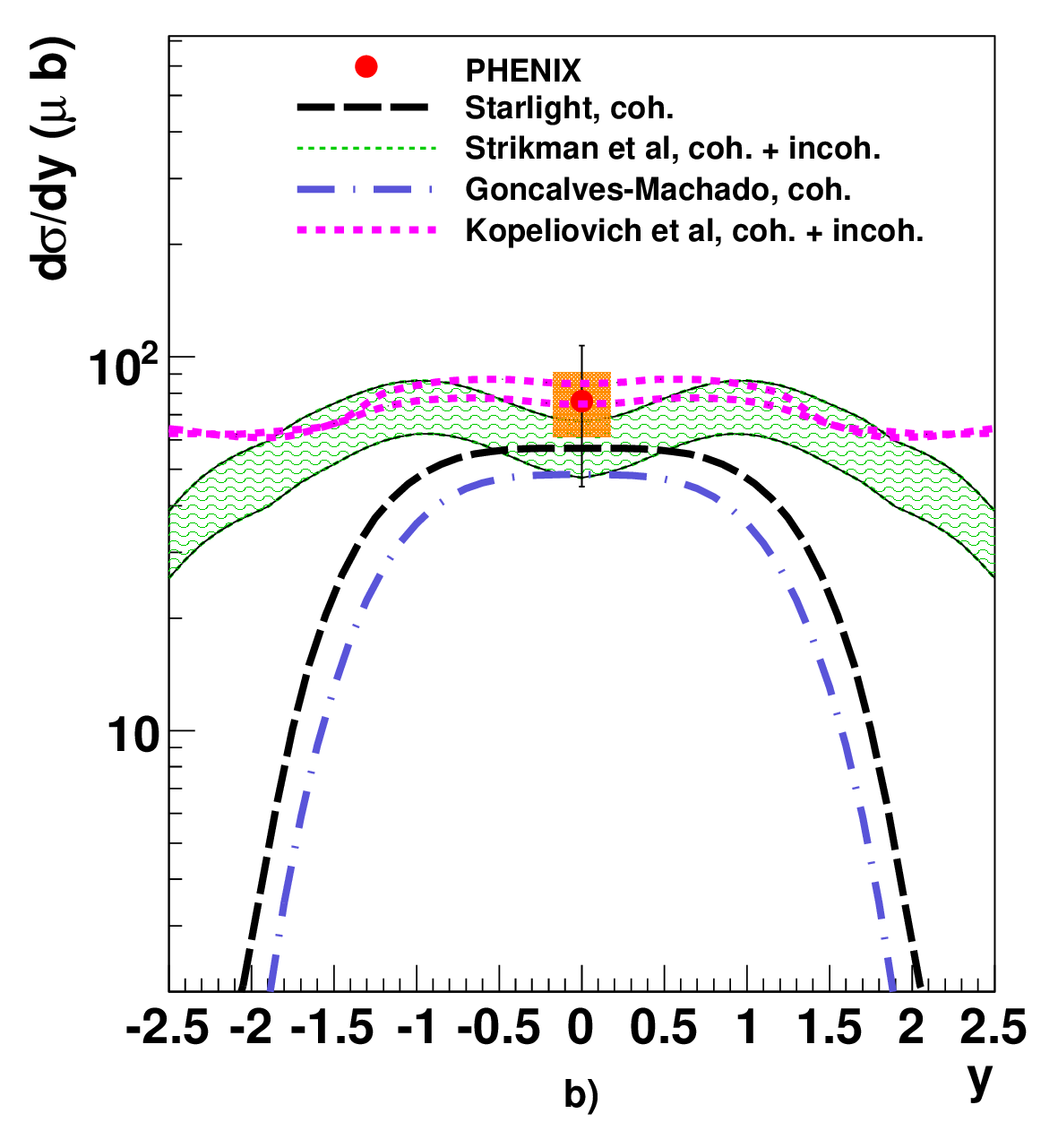}
\end{center}
\caption{Measured cross section of $J/\psi + Xn$ production at midrapidity in UPC
Au+Au collisions at $\sqrt{s_{_{NN}}}$~=~200~GeV.  The error bars (boxes) show
the statistical (systematical) uncertainties.
When available, the theoretical calculations for the coherent
and incoherent components are shown separately in (a), and summed up in
(b).} 
\label{fig:dNdy_vs_model} 
\end{figure}

The final $J/\psi$ cross section is in good agreement, within the (still large) statistical errors, with the theoretical 
values computed 
in~\cite{Baltz:2002pp,Strikman:2005ze,Klein:1999qj,Ivanov:2007ms,Nystrand:2004vn,Goncalves:2007qu} 
as shown in Fig.~\ref{fig:dNdy_vs_model}. For details of the comparision see Ref.~\cite{Afanasiev:2009hy}

\begin{footnotesize}

\bibliographystyle{prlsty}
\bibliography{Master}

\end{footnotesize}

\end{document}